\documentclass[aps,pra,amsmath,amssymb,reprint,floatfix,superscriptaddress]{revtex4-1}

\usepackage[utf8]{inputenc}
\usepackage[T1]{fontenc}
\usepackage[english]{babel}
\usepackage[dvipsnames,svgnames,x11names]{xcolor}
\usepackage{mathtools,dsfont,braket}
\usepackage{tensor}
\usepackage{nicefrac}
\usepackage[version=4]{mhchem}
\usepackage{siunitx}

\usepackage{graphicx}
\usepackage{adjustbox}
\usepackage{float}
\usepackage{tabularx}
\usepackage{booktabs}
\usepackage{multirow}
\usepackage{longtable}
\usepackage{arydshln}
\usepackage{scrextend}

\usepackage{dcolumn}
\usepackage{bm}
\usepackage{microtype}
\usepackage{listings,xcolor}
\usepackage{pgfplots}
\usepackage[pdftex,colorlinks]{hyperref}		 
\hypersetup{%
  linkcolor=blue,%
  citecolor=blue,%
  urlcolor=blue,%
  pdftitle={Manuscript - High-accuracy
  Rb\textsubscript{2}\textsuperscript{$+$} interaction potentials
  based on coupled cluster calculations},%
  pdfauthor={Jan Schnabel, Lan Cheng, Andreas K\"ohn}%
}
\usepackage[outdir=./]{epstopdf}
\usepackage{lipsum}
\usepackage{xr}
\def\expN(#1E#2){\ensuremath{#1 \times 10^{#2}}}

\definecolor{uniddblue}{RGB}{0,65,145}   
\definecolor{uninred}{RGB}{255,0,0}      
\definecolor{unigray}{RGB}{22,22,24}     
\definecolor{unidblue}{RGB}{0,143,255}   
\definecolor{unibblue}{RGB}{176,173,5}   
\definecolor{unired}{RGB}{238,77,46}     

\begin{document}

\preprint{Version1}

\title{High-accuracy
  Rb\textsubscript{2}\textsuperscript{$+$} interaction potentials
  based on coupled cluster calculations} 

\author{Jan Schnabel}
\email[]{schnabel@theochem.uni-stuttgart.de}
\affiliation{Institute for Theoretical Chemistry and Center for Integrated
  Quantum Science and Technology, University of Stuttgart,
  70569 Stuttgart, Germany}

\author{Lan Cheng}
\affiliation{Department of Chemistry, The Johns Hopkins University, Baltimore,
  Maryland 21218, United States}
\email{lcheng24@jhu.edu}

\author{Andreas Köhn}
\email[]{koehn@theochem.uni-stuttgart.de}
\affiliation{Institute for Theoretical Chemistry and Center for Integrated
  Quantum Science and Technology, University of Stuttgart,
  70569 Stuttgart, Germany}

\date{\today}

\begin{abstract}
This work discusses a  protocol for constructing highly accurate potential energy curves
(PECs) for the lowest two states of \ce{Rb2+},
i.e. $X\,\tensor*[^2]{\Sigma}{_g^+}$ and $(1)\,\tensor*[^2]{\Sigma}{_u^+}$,
using an additivity scheme based on coupled-cluster  theory. The approach
exploits the findings of our previous work [J. Schnabel, L. Cheng and A. Köhn,
  J. Chem. Phys. 155, 124101 (2021)] to avoid the unphysical
repulsive long-range barrier occurring for symmetric molecular ions when
perturbative estimates of higher-order cluster operators are
employed. Furthermore, care was taken to reproduce the physically correct
exchange splitting of the $X\,\tensor*[^2]{\Sigma}{_g^+}$ and
$(1)\,\tensor*[^2]{\Sigma}{_u^+}$ PECs. The accuracy of our computational
approach is benchmarked for ionization energies of \ce{Rb} and for
spectroscopic constants as well as vibrational levels of the
$a\,\tensor*[^3]{\Sigma}{_u^+}$ triplet state of \ce{Rb2}. We study
high-level correlation contributions, high-level relativistic effects
and inner-shell correlation contributions
and find very good agreement with experimental reference values for the atomic ionization potential and the binding energy of \ce{Rb2} in the $a\,\tensor*[^3]{\Sigma}{_u^+}$ triplet state.
Our final best estimate for the binding energy of the \ce{Rb2+} $X\,\tensor*[^2]{\Sigma}{_g^+}$
state including zero-point vibrational contributions is $D_0 = 6179\,\mathrm{cm}^{-1}$ with an estimated error bound of $\mathcal{O}(\pm 30\,\mathrm{cm}^{-1})$.
This value is smaller than the experimentally inferred lower bond of $D_0\ge 6307.5\,\mathrm{cm}^{-1}$ [Bellos \emph{et al.}, Phys. Rev. A 87, 012508 (2013)] and will require further investigation. 
For the $(1)\,\tensor*[^2]{\Sigma}{_u^+}$ state a shallow potential with $D_0 = 78.4\,\mathrm{cm}^{-1}$ and an
error bound of $\pm 9\,\mathrm{cm}^{-1}$ is computed.
\end{abstract}


\maketitle 


\section{\label{sec:Intro}Introduction}

The recent experimental
progress~\cite{Schmid2018,Kleinbach2018,Engel2018,Dieterle2020,Dieterle2021,Veit2021}
towards entering the ultracold ($T < 1\,\mathrm{mK}$) quantum regime of hybrid
ion-atom systems is naturally entangled with the need for accurate theoretical models based
on first principles. 
While the cold
regime ($T>1\,\mathrm{mK}$) of ion-atom collisions is essentially classical,
the ultracold domain allows for $s$-wave collisions and thus for reaching the
pure quantum regime. The realization is still a non-trivial task for hybrid
ion-atom systems due to more stringent temperature requirements, in particular
for \ce{Rb2+}~\cite{RevModPhys.91.035001}. Ultracold ion-atom systems provide
a rich platform allowing for the discovery of novel phenomena and
applications~\cite{Kleinbach2018}. Among others, those may reach from
precision measurements of ion-atom collision parameters and associated
molecular potentials~\cite{Cote2000,Idziaszek2009,Schmid2018} to ultracold
state-resolved quantum chemistry~\cite{Jachymski2013} and to the ultimate goal of
realizing strongly coupled charge-neutral polaron
systems~\cite{Casteels2011,Astrakharchik2021,Bissbort2013}. For a
comprehensive overview on both the theoretical and experimental
state-of-the-art research on cold hybrid ion-atom systems see, e.g.,
Refs.~\cite{Cote2016,RevModPhys.91.035001}.

The successful experimental realization of an ultracold ion-atom system has
been demonstrated in Ref.~\cite{Schmid2018} for \ce{Li2+}. The idea of this
class of experiments starts with implanting an ionic impurity into a
Bose-Einstein condensate through a single precursor Rydberg atom followed by
subsequent ionization to initiate collisions with the ionic core and the
ground state host gas atoms. 

For \ce{Rb2+}, initial experiments date back to the 1960s to 1980s and have been
performed in \ce{Rb} vapor with densities of $2.69\cdot
10^{19}\,\mathrm{atoms}\,\mathrm{cm}^{-3}$~\cite{Lee1965}. Corresponding
measurements range from associative photoionization~\cite{Lee1965,Shafi1972}
to rough estimates based on the analysis of charge exchange cross
sections~\cite{Olsen1969} to multiphoton ionization of \ce{Rb2} and subsequent
dissociation of dimer ions by one or more additional
photons~\cite{Wagner1985}. More recent
experiments~\cite{Schmid2018,Kleinbach2018,Engel2018,Dieterle2020,Dieterle2021,Veit2021}
used the same technique as described for \ce{Li2+}. In this way, the
diffuse transport dynamics of the impurity through the BEC and ion-atom-atom
three-body recombination could be
observed~\cite{Dieterle2020,Dieterle2021,Jachymski2020}. In
Ref.~\cite{Dieterle2020} it was even possible to estimate the binding energies
of some threshold bound states. These experiments may thus offer a way to
probe chemical reaction channels at the quantum level. This so-called
state-to-state chemistry will require resolving the quantized molecular energy
levels both theoretically and experimentally.

Recent theoretical investigations of \ce{X2+} systems (with \ce{X}$=$\ce{Li},
\ce{Na}, \ce{K}, \ce{Rb}) can be found, e.g., in
Refs.~\cite{Magnier1999,Magnier2003,Jraij2003,Aymar2003,Berriche2013,Musial2015,Bewicz2017,Skupin2017}. The
equation-of-motion coupled-cluster method for electron attached states  
with single and double replacements (EA-EOM-CCSD) or
even including triple replacements (EA-EOM-CCSDT) and including
scalar-relativistic effects via the Douglas-Kroll-Hess method has been used
for calculations of \ce{Li2+}, \ce{Na2+}, and
\ce{K2+}~\cite{Musial2015,Bewicz2017,Skupin2017}. The authors reported
satisfactory agreement with available experimental results. As perhaps the
only work aiming at highest possible accuracy, Schmid \emph{et al.} reported
\emph{ab-initio} calculations for the \ce{Li2+} interaction potential based on
an additivity scheme with coupled-cluster (CC) computations and large basis
sets. These findings accompany the respective experimental work in
Ref.~\cite{Schmid2018}, which were for the first time precise enough to
predict reasonable bounds for the \ce{Li}+\ce{Li+} scattering length.

With the present work we attempt to provide a general route for obtaining
high-accuracy interaction potentials for the alkaline \ce{X2+} systems based
on an additivity scheme using CC theory. The approach is inspired by the HEAT
(High accuracy Extrapolated Ab-initio Thermochemistry)
protocol~\cite{HEAT2004,HEAT2006,HEAT2008}, which allows for computing binding
energies with uncertainties on the order of $<1\,\mathrm{kJ/mol}$, which translates into
$<84\,\mathrm{cm}^{-1}$. By including our latest
findings~\cite{Schnabel2021,Schnabel2021_PhD} to circumvent the wrong
asymptotic behavior of some of the CC approximations required in this scheme, we report \ce{Rb2+} PECs
with so far unprecedented precision. The binding energy derived from the computed
$X\,\tensor*[^2]{\Sigma}{_g^+}$ potential can be compared to the experimentally
derived~\cite{Bellos2013} lower bound for $D_0^{\text{exp}}$. This experiment
aimed at measuring an upper bound to the ionization energy of
${}^{85}\ce{Rb2}$ formed via photoassociation, which also provides a lower
bound to the dissociation energy of \ce{Rb2+}. 
\begin{figure}[tb]
  \centering
  \includegraphics[width=\columnwidth]{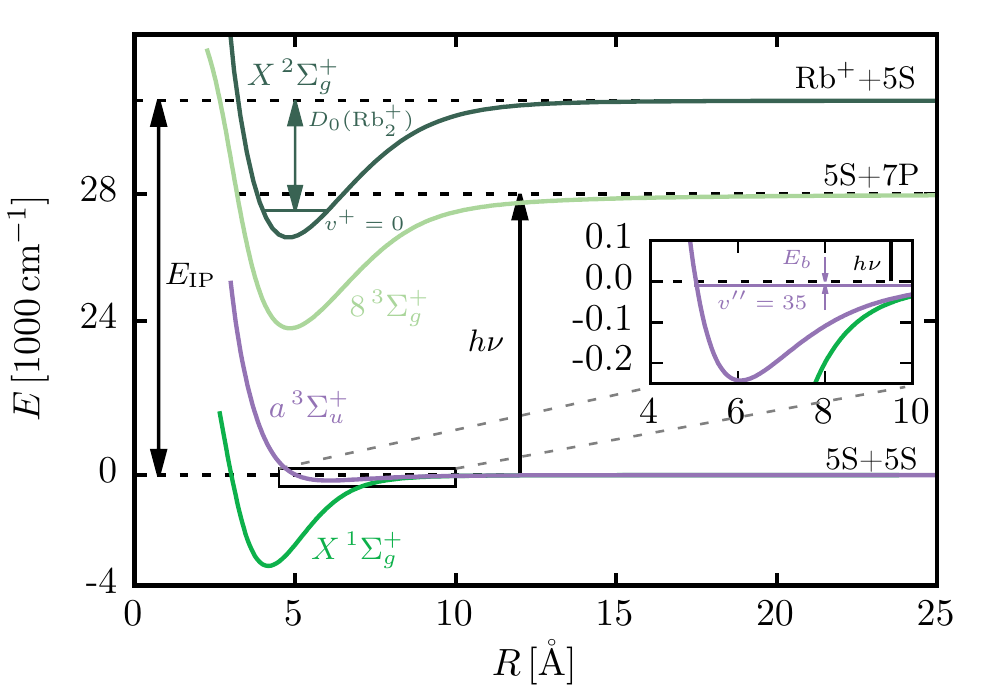}
  \caption{\label{fig:Rb2-PECs_schematic-overview}Schematic overview on
    relevant potential energy curves of \ce{Rb2} and \ce{Rb2+} to obtain a
    lower bound for $D_0(\ce{Rb2+})$ based on the experimental approach
    reported in Ref.~\cite{Bellos2013}. This requires the atomic \ce{Rb}
    ionization energy $E_{\text{IP}}$, the binding energy $E_b$ (see inset) of
    the $v^{\prime\prime}=35$ vibrational level of the
    $a\,\tensor*[^3]{\Sigma}{_u^+}$ state of \ce{Rb2} (initial state for
    photoexcitation) as well as the photon energy $h\nu$ required for the
    onset of autoionization (via a ``superexcited'' level of the
    $8\,\tensor*[^3]{\Sigma}{_g^+}$ state of \ce{Rb2}). \emph{Inspired by a
      figure of M. A. Bellos et al., Phys. Rev. A \textbf{87}, 012508
      (2013).}}
\end{figure}
The bounds were measured by the onset of autoionization of the excited states
of ${}^{85}\ce{Rb2}$ below the 5S$+$7P atomic limit. As illustrated in
Fig.~\ref{fig:Rb2-PECs_schematic-overview}, $D_0^{\text{exp}}$ of \ce{Rb2+}
can be derived given the atomic ionization potential $E_{\text{IP}}$, the
binding energy $E_b[\ce{Rb2}(v^{\prime\prime}=35)]$ of the initial state and
the applied photon energy $h\nu$, which causes the onset of autoionization.

Besides computing spectroscopic constants, we provide analytical
representations of our precise \emph{ab-initio} potentials, which may serve as
foundation to facilitate more sophisticated scattering calculations for
more quantitative statements on the conditions needed to identify effects for
\ce{Rb}$+$\ce{Rb+} collisions that go beyond the Langevin regime. Conventional
approaches~\cite{Jachymski2013,Jachymski2014,Jachymski2020} for studying these
collision processes are merely based on model potentials.

This paper is organized as follows. Section~\ref{sec:ComputationalAspects}
introduces several computational aspects of this work, including the
additivity scheme proposed to potentially reach high accuracy. We gauge the
expected accuracy and limitations of our approach in Sec.~\ref{sec:Benchmark}
in terms of benchmark calculations for ionization energies of \ce{Rb} and
spectroscopic constants as well as vibrational levels of the
$a\,\tensor*[^3]{\Sigma}{_u^+}$ triplet ground state of \ce{Rb2}. Hereafter we
investigate the prospects for obtaining highly accurate ground state PECs of
the \ce{Rb2+} system in Sec.~\ref{sec:HighAccuracyPECsRb2p}. Therefore, we
provide a protocol for constructing physically meaningful interaction
potentials based on CC calculations. By thoroughly investigating
basis set effects and high-level correlation contributions, we infer
reasonable error bars of our computational approach and demonstrate that the
latter are less important to closely reproduce experimental findings.
We provide spectroscopic constants
and explore the rovibrational structure of the $X\,\tensor*[^2]{\Sigma}{_g^+}$
and $(1)\,\tensor*[^2]{\Sigma}{_u^+}$ states of \ce{Rb2+}. Finally, we
summarize the main findings of this work in Sec.~\ref{sec:Conclusion}.

\section{\label{sec:ComputationalAspects}Computational Aspects}

High-accuracy quantum-chemical calculations of atomic and molecular energies
often rely on additivity
schemes~\cite{Feller93,Helgaker97,Martin99,HEAT2004,Schuurman04,Feller08}. Our
approach follows the HEAT protocol~\cite{HEAT2004,HEAT2006,HEAT2008}, 
where we assume that the total
electronic energy $E$ can be calculated via
\begin{align} 
  \label{eq:HEATorig}
  E &= E_{\mathrm{HF}}^\infty + \Delta E_{\mathrm{CCSD(T)}}^\infty + \Delta
  E_{\mathrm{HLC}} + \Delta E_{\mathrm{HLR}}\,. 
\end{align}
Herein, $E_{\mathrm{HF}}^\infty$ and $\Delta E_{\mathrm{CCSD(T)}}^\infty$ are
the Hartree-Fock energy and the CCSD(T) correlation energy, respectively,
at the complete basis set (CBS) limit. Correlation effects beyond
CCSD(T) are denoted as high-level correlation (HLC) contributions $\Delta
E_{\text{HLC}}$ and high-level relativistic (HLR) effects are labeled by $\Delta
E_{\text{HLR}}$. 

The Hartree-Fock and coupled-cluster calculations are performed using either the
small-core effective core potential (scECP) ECP28MDF from
Ref.~\cite{SmallECPStoll}, where only the 4$s^2$4$p^6$5$s^1$ electrons of
\ce{Rb} are treated explicitly and all the others are modelled via a
scalar-relativistic pseudopotential (PP) or using the all-electron spin-free
exact two-component theory in its one-electron variant
(SFX2C-1e)~\cite{Dyall01,Liu09} to treat scalar-relativistic effects. We use
the restricted open-shell (ROHF) approach
for the Hartree-Fock part and for generating the orbitals for the subsequent
single-reference CC calculations. For the latter, we applied an unrestricted
spin-orbital formalism in its singles and doubles variant augmented with a
noniterative triples method based on a ROHF reference, the ROHF-CCSD(T)
method~\cite{Raghavachari89,Bartlett90,Hampel92,Watts92}. 

The HLC contributions considered in this work include the full triples
correction obtained as the difference between full CC singles doubles and
triples (CCSDT)~\cite{Noga87a,Noga1988,Scuseria88,Watts-CCSDT} and CCSD(T)
results using smaller basis sets (i.e., e.g., triple- and quadruple-$\zeta$
quality) and the quadruples correction obtained as the difference between CC
singles doubles triples augmented with noniterative quadruples
[CCSDT(Q)]~\cite{Kucharski1989,Bomble05,Kallay05}
and CCSDT results (at the triple-$\zeta$ level);
cf. Sec.~\ref{sec:Benchmark}. Here we used the CCSDT(Q)/B variant for ROHF
references~\cite{Kallay08}. The same convergence thresholds for
the HF and CC calculations as described in our previous work on the \ce{Rb2+}
system~\cite{Schnabel2021} were employed.

The aug-cc-pCV$n$Z-PP and aug-cc-pwCV$n$Z-PP ($n=3,4,5$)
correlation-consistent basis sets for alkali and alkaline earth atoms,
designed for the ECP28MDF pseudopotential~\cite{GrantHillBasis}, have been used. Furthermore, motivated by the
promising results obtained in our recent work on
\ce{Rb3}~\cite{Schnabel2021_PRA}, we constructed a large
[17s14p9d7f6g5h4i] uncontracted even-tempered basis set ($\equiv$ UET17) based on the
given valence basis set coming with the ECP28MDF pseudopotential in
Ref.~\cite{SmallECPStoll}; see supplementary material~\cite{Supplementary} for details. 
To obtain a better estimate on the Hartree-Fock limit, we combined an
uncontracted and extended version of the aug-cc-pCV5Z-PP basis set for s-,
p-, d- and f-exponents with diffusely augmented g-, h-, and i-functions of the
UET17 basis; cf. supplementary material~\cite{Supplementary}
for technical details and Tab.~S.II for the respective
exponents. We will call this basis ``reference'' in the following.

For the computations with the SFX2C-1e approach a different type of basis set is required.
As the corresponding
aug-cc-pwCV$n$Z-X2C basis sets are only available up to
$n=4$~\cite{GrantHillBasis},  we used the s-, p- and d-type primitive
functions of the uncontracted ANO-RCC (23s,19p,11d) set~\cite{Roos2004} and augmented them
with f-, g-, h-, and i-type functions of our UET17 basis set, which were further complemented by
four steeper functions to be able to describe core-valence correlation effects from the M shell (3s3p3d), 
see Tab.~S.III of the supplementary material~\cite{Supplementary}. These basis sets
will be called UANO-$n$ with $n=3,4,5,6$ in this work. To finally check
basis set saturation in the Hartree-Fock contribution, we replaced the s-, p-, and d-type functions by the decontracted
basis functions from the aug-cc-pwCVQX-X2C basis and will call this basis ``reference (ae)''.
The h- and i-type functions were skipped  in this case to keep the size of the calculations
at a manageable level. These higher-order angular momentum functions are not expected to contribute significantly at the
Hartree Fock level.
See Tab.~S.IV  of the supplementary material~\cite{Supplementary}.

In detail, we used the following protocol for evaluating the theoretical best estimates of the total energy according 
to Eq.\ \eqref{eq:HEATorig}:
The Hartree-Fock CBS limit is obtained by a separate computation with the `reference' basis sets, as described above:
\begin{align}
  \label{CBSHF}
  E_{\text{HF}}^\infty &\approx E_{\text{HF}}(\text{`reference' basis})\,,
\end{align}
The correlation energy at the CBS limit is computed as
\begin{align}
  \label{BSECCSDptCorrEnerg}
  \Delta E_{\text{CCSD(T)}}^\infty 
  &\approx E_{\text{singles}}(n=n_{\text{max}}) + E_{\text{pair}}^\infty + E_{\text{(T)}}^\infty\,.
\end{align}
Here, $E_{\text{singles}}$ is the
CCSD energy contribution resulting from the non-fulfillment of the Brillouin
condition for the ROHF orbitals. We found it advantageous not to extrapolate 
this contribution along with the pair and triples energies as it has a distinctly different
convergence behavior and saturates quickly with increasing cardinal number. Thus, we prefer to
take the value for the highest available basis set result ($n=6$).
The pair energy ($E_{\text{pair}}^\infty$) and the noniterative triples
($E_{\text{(T)}}^\infty$) contributions are extrapolated to their respective
CBS limit using the conventional two-point $n^{-3}$
formula~\cite{BSEStandard1,BSEStandard2}. The same procedure was followed for ECP and SFX2c-1e computations.
In the latter case, we either correlated only the 4s4p and 5s shell (the electrons explicitly treated
in the ECP approach) or we also included the 3s3p3d shell (M shell) into the correlation treatment to
investigate further core valence correlation effects.

The HLC contributions are computed via
\begin{subequations}
  \begin{eqnarray}
    \Delta E_{\text{T}}^{\text{TZ/QZ}} &=& E_{\text{CCSDT}}^{\text{TZ/QZ}} -
    E_{\text{CCSD(T)}}^{\text{TZ/QZ}}\,, \label{BSECCSDTCorrEnergT}\\
    \Delta
    E_{\text{(Q)}}^{\text{TZ}} &=& E_{\text{CCSDT(Q)}}^{\text{TZ}}-E_{\text{CCSDT}}^{\text{TZ}}\,,\label{BSECCSDTCorrEnergQ}
  \end{eqnarray}
\end{subequations}
where we relied on ECP based calculations and the aug-cc-pCV$n$Z-PP ($n$ = T, Q) basis sets.
As indicated by the superscript ``TZ/QZ'', the increment for the CCSDT contribution is
obtained by extrapolating the correlation energies
with the aug-cc-pCVTZ-PP and aug-cc-pCVQZ-PP basis sets using the two-point $n^{-3}$ formula. 
The CCSDT(Q) computations used the aug-cc-pCVTZ-PP basis. 

The considered high-level relativistic effects are relevant as corrections for the SFX2C-1e approach only.
They consist of the two-electron picture-change (2e-pc)
corrections, spin-orbit (SO) corrections, and the contributions from the Breit term. 
In addition there are effects from quantum electrodynamics (QED), which we 
did not attempt to compute here.
The 2e-pc correction
is obtained as the difference between the spin-free Dirac-Coulomb (SFDC)\cite{Dyall:JCP1994-2118} and
SFX2C-1e results.  The SO correction together with the contribution from the Breit term is
calculated as the difference between the spin-orbit X2C scheme with atomic
mean-field spin-orbit integrals (X2CAMF) based on the Dirac-Coulomb-Breit Hamiltonian\cite{Liu:JCP2018-144108,Zhang:JPCA-ipr} 
and the SFX2C-1e scheme.
Calculations of these contributions have been carried out at the
CCSD(T) level using the uncontracted ANO-RCC basis set, employing the frozen-core approximation for the 1s2s2p3s3p3d shells and 
deleting virtual orbitals higher than 1000 $E_\text{h}$. 

We note that it is less straightforward to incorporate $\Delta
E_{\text{HLR}}$ into ECP calculations. This is due to the fact that the
ECP28MDF pseudopotential already contains a two-component spin-orbit coupled
part with corresponding parameters adjusted to valence energies obtained at
all-electron multiconfiguration Dirac-Coulomb-Hartree-Fock (DC-HF) level of
theory, which includes relativistic effects at a four-component level of
theory~\cite{SmallECPStoll}. Therefore, one cannot use the difference
between SFX2C-1e and high-level relativistic effects and ECP results  
as the correction to ECP. Furthermore, the quality of the basis
sets used in the SFX2C-1e calculations are not exactly the same as those 
for the ECP calculations, so taking the
difference introduces additional errors.

All ROHF-CCSD(T) calculations have been performed with the
\textsc{Molpro} 2019.2 program
package~\cite{MolproReview2012,MolproReview2020,MOLPRO2019-2}, the
 CCSDT calculations have been carried out using the
\textsc{cfour} program
package~\cite{cfour,Matthews2020CFOUR,Watts92,Cheng11b,Liu:JCP2018-034106}, and all CCSDT(Q)
energies were computed with the \textsc{Mrcc} program
suite~\cite{MRCCProg,generalCCKallay,Kallay05,Kallay2020}.

\section{\label{sec:Benchmark}Benchmark Calculations}
In the following we benchmark the protocol described in Sec.~\ref{sec:ComputationalAspects} to assess both its expected accuracy and its
limitations based on experimental reference values for \ce{Rb} and \ce{Rb2}.

\subsection{\label{subsec:Rb-IP}Rb ionization potential}

Calculations of the ionization potential (IP) of atomic \ce{Rb} are listed
in Tab.~\ref{tab:RbIP}.
\begin{table*}[tb]
  \centering
  \caption{\label{tab:RbIP}Results for the \ce{Rb} ionization energy
    $E_{\mathrm{IP}}$ (in $\mathrm{cm}^{-1}$). The 4s4p and 5s shells are correlated unless noted otherwise.
    The experimental value is
    $E_{\mathrm{IP}}^{\mathrm{exp.}}=33690.81\,\mathrm{cm}^{-1}$~\cite{NISTRbAtomic,Lorenzen1983}.}
  \begin{adjustbox}{max width=\textwidth}
    \begin{tabular}{lSSSS}
      \toprule\toprule
      & \multicolumn{2}{c}{ECP} & \multicolumn{2}{c}{SFX2C-1e} \\
      \cmidrule(r){2-3}       \cmidrule(r){4-5} 
      &\multicolumn{1}{c}{apCV$n$Z-PP} & \multicolumn{1}{c}{UET17} &
      \multicolumn{1}{c}{UANO} &\multicolumn{1}{c}{UANO (M shell)\footnote{3s3p3d shell included in correlation treatment.}}\\
      \midrule
      {$n=4$} & 33550.89 & 33644.35 & 33561.48 & 33639.48 \\
      {$n=5$} & 33606.14 & 33664.46 & 33581.71 & 33660.87 \\
      {$n=6$} & {--}     & 33671.71 & 33589.01 & 33668.26 \\
      \midrule
      CBS (corr.)\footnote{Extrapolation of correlation energy only, Hartree-Fock energy from largest $n$.}            
                       & 33665.87 & 33681.74 & 33599.11 & 33678.50 \\
      CBS (HF+corr.)\footnote{Including a correction for Hartree-Fock limit using the `reference' basis sets.}
                       & {--}     & 33675.96 & 33601.49 & 33680.89 \\
      \midrule\midrule
      {$\Delta E_{\text{T}}^{\text{TZ/QZ}}$}    &  +2.40 & [+2.40] & [+2.40] & [+2.40] \\
      {$\Delta E_{\text{(Q)}}^{\text{TZ}}$}     &  +7.63 & [+7.63] & [+7.63] & [+7.63] \\
      {$\Delta E_{\text{(Q)}}^{\text{TZ/QZ}}$}  & +10.72 &[+10.72] &[+10.72] &[+10.72] \\
      \midrule
      {$+\Delta E_{\text{HLR}}$} & {--} & {--} & 17.75 & 17.75 \\
      \midrule\midrule
      {TBE}\footnote{Theoretical best estimate (see text). }   & 33675.90 & 33685.99 & 33629.57 & 33708.96 \\
      {TBE(ext)\footnote{Theoretical best estimate using $\Delta E_{\text{(Q)}}^{\text{TZ/QZ}}$ instead of $\Delta E_{\text{(Q)}}^{\text{TZ}}$}} 
                                                               & 33678.99 & 33689.08 & 33632.66 & 33712.05 \\
      \bottomrule\bottomrule
    \end{tabular}
  \end{adjustbox}
\end{table*}
Comparison to the results using the aug-cc-pCV$n$Z-PP basis shows that the UET17 indeed leads to a much tighter convergence 
of the correlation energy. The gap between the $n=6$ result and the estimated CBS limit of the IP is only $\sim$10$\,\mathrm{cm}^{-1}$
and this observation also holds for the all-electron calculations with the UANO basis. We may therefore estimate the 
uncertainty of this result to be at most half of this difference. It is also evident from the two last columns of Tab.~\ref{tab:RbIP} 
that there is a significant contribution from the M shell correlation of nearly 80$\,\mathrm{cm}^{-1}$. Interestingly, the ECP result
(that only implicitly includes the effect of inner shells) is rather close to the SFX2C-1e result \textit{with} explicit M shell correlation, see line `CBS' in Tab.\ \ref{tab:RbIP}.
As the effect of core-correlation will be much smaller in the other cases where we mainly look at binding energies, we did not
go into further detail regarding this observation. Clearly, this points at the limits of the accuracy of the ECP approximation. 

Moving to a larger basis for the Hartree-Fock contribution leads to corrections of the IP by 6 and 2$\,\mathrm{cm}^{-1}$ for the ECP approach and the all-electron calculations, respectively. As a conservative estimate of the residual uncertainty for the Hartree-Fock contribution we thus take 5$\,\mathrm{cm}^{-1}$. The high-level correlation effects have been evaluated for the ECP/aug-cc-pCV$n$Z-PP computations only. The overall result is a contribution of 10$\,\mathrm{cm}^{-1}$. For the atoms, it was also feasible to carry out the CCSDT(Q) calculations with
a quadruple-$\zeta$ quality basis set, which allowed a basis set extrapolation for this quantity as well, leading to another 3$\,\mathrm{cm}^{-1}$. Overall, the contribution of $\Delta E_\text{HLC}$ is not very large and we will in the following use its total value as an estimate of the overall uncertainty in the correlation energy due to higher-order correlation (for the case of the IP of Rb: 10$\,\mathrm{cm}^{-1}$).

Finally, we consider the high-level relativistic (HLR) corrections beyond SFX2C-1e, $\Delta E_\text{HLR}$.  
The two-electron picture change correction gives $5.76\,\mathrm{cm}^{-1}$ and the inclusion of spin-orbit coupling and the Breit term contributes $2.01\,\mathrm{cm}^{-1}$ further.  
In addition, a QED correction of
$9.98\,\mathrm{cm}^{-1}$ has been taken from Ref.~\cite{Koziol18}.
These three contributions add up to $\Delta
E_{\text{HLR}}=17.75\,\mathrm{cm}^{-1}$ as given in Tab.~\ref{tab:RbIP}. The uncertainty in this correction may be
conservatively estimated as roughly half its amount (10$\,\mathrm{cm}^{-1}$); 
we will later find much smaller relativistic corrections in the calculation of binding energies, which will then
allow us to gauge the uncertainty by the total size of the high-level relativistic correction.

The final best estimates may be compared to the experimental value of
$E_{\text{IP}}^{\text{exp}}=33690.81\pm
0.01\,\mathrm{cm}^{-1}$~\cite{NISTRbAtomic,Lorenzen1983}. We find a deceptively good coincidence
of the ECP based result, in particular, if the extrapolated CCSDT(Q) results are taken into account. However, the 
error estimates discussed so far add up to approximately $\pm 30\,\mathrm{cm}^{-1}$. Using this error bound, the 
overall X2C-based theoretical best estimate of 33709$\,\mathrm{cm}^{-1}$ is also in good agreement with the experimental
value. The core-correlation from the M shell seems to be important for quantitative predictions and will be 
closely analyzed for its impact on the binding energies. It is most likely dominated by the 3d subshell, 
which is energetically significantly higher than the 3s and 3p subshells.  The correlation from lower-lying shells (K and L) is much
less important, as exploratory computations show (although with not fully sufficient basis sets, as even steeper
polarization function were needed). Using the UANO-6 basis, correlation of the L shell reduces the IP by 3$\,\mathrm{cm}^{-1}$
and of the K shell by further 1$\,\mathrm{cm}^{-1}$, see supplementary
material~\cite{Supplementary}.

In summary, the results for the ionization energy are very promising in terms of obtaining 
first-principle predictions with an accuracy definitely better than 1 kJ/mol ($84\,\mathrm{cm}^{-1}$).

\subsection{\label{subsec:Rb2}Rb\textsubscript{2} -- the $\mathrm{a}\,\tensor*[^3]{\Sigma}{_u^+}$
  state}

Next, we gauge the accuracy of our computational approach  for molecular
calculations and compute the spectroscopic constants $D_e$ and $R_e$, i.e. the depth of the interaction
potential and its equilibrium distance, for the lowest triplet
state $a\,\tensor*[^3]{\Sigma}{_u^+}$ of \ce{Rb2}. This state plays a fundamental 
role in photoassociation processes to produce ultracold \ce{Rb2}
molecules~\cite{Stwalley1999,Jones2006,Ulmanis2012,Strauss2010,Bellos2011,Bellos2013}
and is quite challenging for computational studies  due to the shallow nature of the potential, as displayed in
Fig.~\ref{fig:Rb2-PECs_schematic-overview}. 
Table~\ref{tab:V2BasisDepend} lists the results in a scheme analogous to that used for the IP of atomic Rb.
\begin{table*}[tb]
  \centering
  \caption{\label{tab:V2BasisDepend}
    Computed binding energies $D_e$ and equilibrium bond distances of
    the $a\,\tensor*[^3]{\Sigma}{_u^+}$ state of \ce{Rb2}.
    The experimental values are
    $D_e^{\mathrm{exp}}=241.5045\,\mathrm{cm}^{-1}$ and
    $R_e^{\mathrm{exp}}=\SI{6.0650}{\angstrom}$.\cite{Guan2013}.
    }
  \begin{adjustbox}{max width=\textwidth}
    \begin{tabular}{lSSSSSS}
      \toprule\toprule
      & \multicolumn{2}{c}{ECP} & \multicolumn{2}{c}{SFX2C-1e} & \multicolumn{2}{c}{SFX2C-1e (+ M shell)\footnote{3s3p3d shell included in correlation treatment.}} \\
       \cmidrule(r){2-3} \cmidrule(r){4-5} \cmidrule(r){6-7}
      & {$D_e\,[\mathrm{cm}^{-1}]$} & {$R_e\,[\si{\angstrom}]$} &
        {$D_e\,[\mathrm{cm}^{-1}]$} & {$R_e\,[\si{\angstrom}]$} &
        {$D_e\,[\mathrm{cm}^{-1}]$} & {$R_e\,[\si{\angstrom}]$}  \\
          \midrule
          $n=4$ & 239.08 & 6.081 & 249.20 & 6.068 & 249.77 & 6.058 \\
          $n=5$ & 241.62 & 6.069 & 252.94 & 6.046 & 253.90 & 6.033 \\
          $n=6$ & 241.90 & 6.066 & 253.22 & 6.044 & 254.22 & 6.030 \\
          \midrule
          CBS (corr.)\footnote{Extrapolation of correlation energy only, Hartree-Fock energy from largest $n$.}     & 242.28 & 6.062 & 253.59 & 6.040 & 254.64 & 6.026 \\
          CBS (HF+corr.)\footnote{Including a correction for Hartree-Fock limit using the `reference' basis sets.} & 239.50 & 6.067 & 246.31 & 6.057 & 247.31 & 6.044 \\
          \midrule
          $+\Delta E_{\text{T}}^{\text{TZ/QZ}}$       & 237.58 & 6.067 & 244.39 & 6.058 & 245.39 & 6.044 \\
          $+\Delta E_{\text{(Q)}}^{\text{TZ}}$ = TBE\footnote{Theoretical best estimate (see text). } & 243.25 & 6.058 & 250.09 & 6.049 & 251.11 & 6.036 \\
          \midrule
          $\Delta E_\text{HLR}$ &    &   & -0.30 &   &  & \\
          \bottomrule\bottomrule
    \end{tabular}
  \end{adjustbox}
\end{table*}

Interestingly, in comparison to the known experimental reference, the ECP computations deliver the closest result. 
This close coincidence is largely lucky in view of the estimates for the uncertainty of these values. 
The uncertainty in the extrapolation is rather small in this case, the results for the CBS limit deviate by less than 0.5~cm$^{-1}$ from the $n=6$ value in all cases and the impact on the bond distance is less than \SI{0.005}{\angstrom}. The effect of core correlation is also very small, the results of the two sets of X2C-1e computations without and with correlation of the M shell differ by only 1~cm$^{-1}$.
A slightly larger correction is imposed by the Hartree-Fock correction, which reduces the binding energy by nearly 3~cm$^{-1}$ for the ECP calculations and 7~cm$^{-1}$ for the all-electron calculations. 
The high-level corrections for full three-electron clusters and four-electron clusters work in opposite directions, giving corrections of $-2$~cm$^{-1}$ and $+6$~cm$^{-1}$, respectively. 
High-level relativistic corrections turn out to have very little impact on the binding energy, they reduce the binding energy by 0.30~cm$^{-1}$ and thus do not close the small gap between the ECP and the all-electron results. 
In view of the smallness of this correction (a similar size of correction will be found for \ce{Rb2+}) we decided to not include this value in the total sum for the best estimate but only as indicator for the size of effects due to the approximate treatment of relativistic effects.
The total estimate of the uncertainty of the binding energy, obtained in the same way as before for the IP, adds up to $\pm$12~cm$^{-1}$ and thus also spans the difference between the ECP and all-electron calculations. 
The experimental value lies quite at the border of this range, indicating that the all-electron calculations may have the general tendency of overbinding due to basis set superposition errors. 
The predicted equilibrium bond distance follows the trend for the binding energies: The larger the binding energy the shorter the bond. Again in comparison to the experimental value, the all-electron results give slightly too short bond distances.

For comparison, we performed ECP-based CCSD(T) calculations using the aug-cc-pCV$n$Z-PP basis sets yielding $(D_e,R_e)=(266.9\,\mathrm{cm}^{-1},\SI{6.09}{\angstrom})$, $(255.5\,\mathrm{cm}^{-1},\SI{6.06}{\angstrom})$ and $(245.5\,\mathrm{cm}^{-1},\SI{6.07}{\angstrom})$ for the $n=3, 4, 5$ cardinalities, respectively. 
These computations converge to comparable results, but clearly with a much larger uncertainty for the extrapolation to the CBS limit.

We also assessed the size of the spin-orbit splitting of the triplet state by performing
frozen-core MRCI/ECP28MDF/UET17($n=3$) calculations~\cite{MRCI1,MRCI2,MRCI3}
using the ECP-LS technique for the corresponding small-core ECP. The computations
included seven singlet and seven triplet states of \ce{Rb2}, such that in
total a 28$\times$28 spin-orbit matrix is set up and diagonalized. The resulting
zero-field splittings and energy shifts for the $a\,\tensor*[^3]{\Sigma}{_u^+}$
state are both $\ll 1\,\mathrm{cm}^{-1}$,
which suggests that spin-orbit effects can be neglected in the further
discussion. A very small zero-field splitting could also be observed in the X2CAMF
computations.
 This is in accordance with the results reported in
Ref.~\cite{Allouche2012} for the $a\,\tensor*[^3]{\Sigma}{_u^+}$ state of
\ce{Rb2}, yielding effects in the order of $0.5\,\mathrm{cm}^{-1}$
for $D_e$ when comparing numbers obtained with and without SO effects
included.

The ROHF-CCSD(T)/\-ECP28MDF/\-UET17 approach was also used to compute a larger set of points of the potential energy curve
and the one-dimensional RP-RKHS interpolation method as described in
Refs.~\cite{Ho1996,Hollebeek1999,Ho2000a,Ho2000b,Higgins2000,Unke2017,SoldanPESRb3} was used to obtain an analytic representation. The
respective long-range coefficients required by this procedure were taken from
Ref.~\cite{Strauss2010}, with $C_6=0.227003\cdot
10^8\,\mathrm{cm}^{-1}\si{\angstrom}^6$, $C_8=0.778289\cdot
10^9\,\mathrm{cm}^{-1}\si{\angstrom}^8$ and $C_{10}=0.286887\cdot
10^{11}\,\mathrm{cm}^{-1}\si{\angstrom}^{10}$;
cf. Sec.~\ref{subsec:GenFormX2pPots} for details on the asymptotic form of
interaction potentials. This long-range behavior with three reciprocal power
terms defines the RP-RKHS parameters, yielding $n_{\ce{Rb2}}=3$,
$m_{\ce{Rb2}}=2$ and $s_{\ce{Rb2}}=2$ as well as
$R_a=18.0\,\mathrm{a}_0$.  To further provide a
physically meaningful short-range description, we imposed a correction of the
form
\begin{align}
  \label{VSR}
  V_{\text{SR}}(R) &= \frac{a}{R}\exp(-bR)\,.
\end{align}
This only affects the PEC for $R<\SI{4.0}{\angstrom}$ and is introduced to
correct for inherent artefacts of the RP-RKHS procedure concerning short-range
extrapolation~\cite{Soldan2000}. Thus, it has no impact on $D_e$ and $R_e$ but
gets relevant for high-energy scattering states ($E>15000\,\mathrm{cm}^{-1}$).

Inspired by the approach reported in Ref.~\cite{SoldanPESRb3}, we further
provide an empirically adjusted ``optimal'' PEC for which we scaled and
shifted the CBS \emph{ab-initio} data in order to match the experimentally derived
values $D_e^{\text{exp}}$ and $R_e^{\text{exp}}$,
respectively~\cite{Schnabel2021_PhD}. The resulting PEC can be reproduced
using our programs and data available from Ref.~\cite{SchnabelZenodo2021}. In
the present work, we used this PEC to compute the rovibrational structure of
the $a\,\tensor*[^3]{\Sigma}{_u^+}$ state of \ce{Rb2}. The resulting 41
vibrational levels for $J=0$ are given in Tab.~S.VIII of the supplementary
material~\cite{Supplementary}. We observe excellent agreement with experimentally measured levels.

\section{\label{sec:HighAccuracyPECsRb2p}High-accuracy rubidium ion-atom interaction potentials}

The previous discussion on \ce{Rb} ionization energies and spectroscopic
constants of the $a\,\tensor*[^3]{\Sigma}{_u^+}$ triplet ground state of
\ce{Rb2} demonstrate the capability of our computational approach to predict
energies and potential energy curves to high accuracy. In the following we
investigate the binding energies and the full PECs of the two lowest states of \ce{Rb2+}.

\subsection{\label{subsec:Rb2p-basis-set-effects-HLC}Accurate binding energies of \ce{Rb2+} }

We start out with a discussion of the binding energy and equilibrium distance of the $X\,\tensor*[^2]{\Sigma}{_g^+}$ state.
As already discussed previously\cite{Schnabel2021}, there is in our approach a slight consistency error in the asymptotic region, as the symmetry-adapted mean-field (Hartree-Fock) solution formally dissociates the system into two fragments which are a 50:50 mixture of a cation and a neutral atom.
Thus, the computed energies in the asymptote do not coincide with the sum of those of the atom and the cation from individual mean-field computations. The consistency could only be achieved by allowing for symmetry-broken solutions in the asymptote.
This problem may be fully avoided by special techniques, e.g.\ the electron attachment equation of motion coupled-cluster (EA-EOM-CC) approach \cite{Musial2015,Bewicz2017,Skupin2017}. However, going to higher-order correlation methods, the effect diminishes and eventually does not significantly contribute to the overall uncertainty of the final result. 
At the Hartree-Fock level, using the ECP based approach, the asymptote lies 40~cm$^{-1}$ above the atomic limit, which diminishes to $-4.4$~cm$^{-1}$ at the CCSD(T) level (nearly independent of the basis set size). At the CCSDT and CCSDT(Q) level, the error shrinks to $+1.5$~cm$^{-1}$ and $+0.7$~cm$^{-1}$ respectively. In the SFX2C-1e scheme, the same observations are made, differing by only $0.3$~cm$^{-1}$ from the values quoted above. 
We will thus in the following report energies relative to those computed for the separated ion and atom. The main results are summarized in Tab.~\ref{tab:Rb2pGest}, using the same scheme as discussed before.
\begin{table*}[tb]
  \centering
  \caption{\label{tab:Rb2pGest}
    Computed binding energies $D_e$ and equilibrium bond distances of
    the $X\,\tensor*[^2]{\Sigma}{_g^+}$ state of \ce{Rb2+}.
    }
  \begin{adjustbox}{max width=\textwidth}
    \begin{tabular}{lcccccc}
      \toprule\toprule
      & \multicolumn{2}{c}{ECP} & \multicolumn{2}{c}{SFX2C-1e} & \multicolumn{2}{c}{SFX2C-1e (+ M shell)\footnote{3s3p3d shell included in correlation treatment.}} \\
       \cmidrule(r){2-3} \cmidrule(r){4-5} \cmidrule(r){6-7}
      & {$D_e\,[\mathrm{cm}^{-1}]$} & {$R_e\,[\si{\angstrom}]$} &
        {$D_e\,[\mathrm{cm}^{-1}]$} & {$R_e\,[\si{\angstrom}]$} &
        {$D_e\,[\mathrm{cm}^{-1}]$} & {$R_e\,[\si{\angstrom}]$}  \\
          \midrule
          $n=4$ & 6157.6 & 4.813 & 6171.9 & 4.820 & 6181.4 & 4.806\\
          $n=5$ & 6176.8 & 4.806 & 6198.1 & 4.810 & 6210.5 & 4.795 \\
          $n=6$ & 6181.5 & 4.803 & 6203.8 & 4.807 & 6217.0 & 4.792 \\
          \midrule
          CBS (corr.)\footnote{Extrapolation of correlation energy only, Hartree-Fock energy from largest $n$.}   
                         & 6187.7 & 4.801 & 6210.2 & 4.804 & 6224.5 & 4.788 \\
          CBS (HF+corr.)\footnote{Including a correction for Hartree-Fock limit using the `reference' basis sets.} 
                         & 6180.5 & 4.804 & 6189.6 & 4.810 & 6203.6 & 4.795 \\
          \midrule
          $+\Delta E_{\text{T}}^{\text{TZ/QZ}}$       & 6177.4 & 4.806 & 6186.5 & 4.812 & 6200.5 & 4.797 \\
          $+\Delta E_{\text{(Q)}}^{\text{TZ}}$ = TBE\footnote{Theoretical best estimate (see text). } 
                                                      & 6178.8 & 4.805 & 6187.9 & 4.811 & 6201.9 & 4.796 \\
                                                      \midrule
          $\Delta E_\text{HLR}$ & & & \phantom{0}$-0.4$ & & & \\
          \bottomrule\bottomrule
    \end{tabular}
  \end{adjustbox}
\end{table*}

The CCSD(T) binding energies at the CBS limit differ by approximately 20~cm$^{-1}$ when evaluated from either the ECP or the SFX2C-1e all-electron approach. When correcting for the Hartree-Fock limit, using again large basis sets in both approaches, the difference shrinks to less than 10~cm$^{-1}$. 
Correlating the M shell in the latter approach adds 14~cm$^{-1}$ to the binding energy. 
The difference between ECP and SFX2C-1e results cannot be attributed to remaining high-level relativistic effects. By comparison of SFX2C-1e and SFDC computations, we can estimate a two-electron picture-change effect on the order of $-0.14$~cm$^{-1}$ whereas going to the two-component X2CAMF scheme gives a correction of the dissociation energy by $-0.26$~cm$^{-1}$, resulting in a total $\Delta E_{\mathrm{HLR}}$ of $-0.4$~cm$^{-1}$. The high-level correlation effects are also rather small. Including full triply connected clusters decreases the binding energy by 3 cm$^{-1}$ and the correction from CCSDT(Q) increases it again by approximately half of this value.

Overall we may estimate an uncertainty of $\pm 10$~cm$^{-1}$ for the Hartree-Fock contribution, $\pm 5$~cm$^{-1}$ for the extrapolation of the CCSD(T) CBS limit and $\pm 5$~cm$^{-1}$ for high-level correlation effects. Further uncertainties may be correlation effects from even deeper core shells (should not exceed $\pm 5$~cm$^{-1}$) and further relativistic effects (which are also unlikely to exceed $\pm 5$~cm$^{-1}$). Our best estimate for the dissociation energy of \ce{Rb2+} in the $X\,\tensor*[^2]{\Sigma}{_g^+}$ state is thus 6202$\pm$30~cm$^{-1}$.

The predicted equilibrium distance changes by less than \SI{0.02}{\angstrom} among all calculations, except for those with the smallest basis set, see Tab.~\ref{tab:Rb2pGest}. Correlation of the M shell leads to a small contraction of the distance by \SI{0.01}{\angstrom} and our best estimate is thus \SI{4.796\pm 0.010}{\angstrom}.

The antisymmetric $(1)\,\tensor*[^2]{\Sigma}{_u^+}$ state also shows a shallow minimum, which is purely due to induced dipole-charge and dispersion effects. The first conclusion is supported by the fact that the computed Hartree-Fock contribution to the binding energy is around 60~cm$^{-1}$ at the equilibrium distance. The best estimate for the binding energy including correlation effects can be deduced from the numbers in Tab.\ \ref{tab:Rb2pUest}.
\begin{table*}[tb]
  \centering
  \caption{\label{tab:Rb2pUest}
    Computed binding energies $D_e$ and equilibrium bond distances of
    the $(1)\,\tensor*[^2]{\Sigma}{_u^+}$ state of \ce{Rb2+}.
     }
  \begin{adjustbox}{max width=\textwidth}
    \begin{tabular}{lcccccc}
      \toprule\toprule
      & \multicolumn{2}{c}{ECP} & \multicolumn{2}{c}{SFX2C-1e} & \multicolumn{2}{c}{SFX2C-1e (+ M shell)\footnote{3s3p3d shell included in correlation treatment.}} \\
       \cmidrule(r){2-3} \cmidrule(r){4-5} \cmidrule(r){6-7}
      & {$D_e\,[\mathrm{cm}^{-1}]$} & {$R_e\,[\si{\angstrom}]$} &
        {$D_e\,[\mathrm{cm}^{-1}]$} & {$R_e\,[\si{\angstrom}]$} &
        {$D_e\,[\mathrm{cm}^{-1}]$} & {$R_e\,[\si{\angstrom}]$}  \\
          \midrule
          $n=4$ &  82.55 & 12.206 & 83.88 & 12.217 & 83.88 & 12.196 \\
          $n=5$ &  82.64 & 12.202 & 83.95 & 12.212 & 83.95 & 12.191 \\
          $n=6$ &  82.66 & 12.200 & 83.97 & 12.210 & 83.97 & 12.189 \\
          \midrule
          CBS (corr.)\footnote{Extrapolation of correlation energy only, Hartree-Fock energy from largest $n$.}
                         &  82.65 & 12.198 & 83.96 & 12.208 & 83.95 & 12.187 \\
          CBS (HF+corr.)\footnote{Including a correction for Hartree-Fock limit using the `reference' basis sets.} 
                         &  82.51 & 12.205 & 82.91 & 12.216 & 82.90 & 12.195 \\
          \midrule
          $+\Delta E_{\text{T}}^{\text{TZ/QZ}}$       
                         & 77.14 & 12.192 & 77.53 & 12.204 & 77.53 & 12.182 \\
          $+\Delta E_{\text{(Q)}}^{\text{TZ}}$ = TBE\footnote{Theoretical best estimate (see text). } 
                         & 80.04 & 12.173 & 80.43 & 12.185 & 80.43 & 12.163 \\
          \bottomrule\bottomrule
    \end{tabular}
  \end{adjustbox}
\end{table*}
In this case there are only little effects from basis set extrapolation, both for the correlation and the Hartree-Fock contribution. Likewise there is only a small discrepancy of 1~cm$^{-1}$  between ECP-based and SFX2C-1e-based computations and a nearly vanishing effect of core correlation. 
High-level relativistic effects have not been estimated for this case due to technical problems in converging the orbitals, but based on previous experience they are expected to be miniscule.
The largest correction comes from high-level correlation effects, a reduction of the binding energy by nearly 5.5~cm$^{-1}$ from the CCSDT calculations and a slight increase again by 2.9~cm$^{-1}$ from CCSDT(Q).
The final best estimate is thus 80$\pm$9~cm$^{-1}$ for the binding energy. 

The equilibrium distance also shows only very little dependence on the basis set size and the biggest correction comes from the high-level correlation effects, being a shortening by \SI{0.015}{\angstrom}. There is also some effect of M core correlation, which shortens the bond length by \SI{0.02}{\angstrom} in comparison to the computation that keeps this shell uncorrelated. Overall, we may give a best theoretical estimate for $R_e$ of \SI{12.163\pm0.02}{\angstrom}.

\subsection{\label{subsec:GenFormX2pPots}General long-range form of \ce{X2+}
  interaction potentials}

An essential aspect of the recent experimental
works~\cite{Schmid2018,Kleinbach2018,Engel2018,Dieterle2020,Dieterle2021,Veit2021}
towards entering the ultracold domain of ion-atom interactions relies on studying
scattering events. The scattering properties of such collisions are defined by
the long-range form of the respective interaction
potentials~\cite{RevModPhys.91.035001}. For ionic dimers with a single active
electron (i.e., e.g., alkali-metal systems such as \ce{Rb2+}) this long-range
behavior contains the two contributions~\cite{Cote2016}
\begin{align}
  \label{VLR}
  V_{\mathrm{LR}}(R) &= V_{\text{ind/disp}}(R) \pm V_{\text{exch}}(R)\,,
\end{align}
where $V_{\text{ind/disp}}(R)$ describes the leading induction and dispersion
interaction, while $V_{\text{exch}}$ defines the exchange interaction
term. For interactions between an $S$-state atom and an $S$-state ion, the
first term of Eq.~\eqref{VLR} becomes~\cite{RevModPhys.91.035001}
\begin{align}
  \label{Vdisp}
  V_{\text{ind/disp}}(R) &= -\frac{C_4^{\text{ind}}}{R^4} -
  \frac{C_6^{\text{ind}}}{R^6} - \frac{C_6^{\text{disp}}}{R^6} + \cdots\,.
\end{align}
In leading order the interaction is due to the charge $q$ of the ion inducing
an electric dipole moment of the atom, which reflects in the corresponding
induction coefficient
\begin{align}
  \label{C4ind}
  C_4^{\text{ind}} &= \frac{1}{2}q^2\alpha_{\text{d}}\,,
\end{align}
with the static electric dipole polarizability $\alpha_{\text{d}}$ of the
atom. The next higher-order induction term is caused by the interaction
between the charge of the ion and the induced electric quadrupole moment of
the atom, described by the respective coefficient 
\begin{align}
  \label{C6ind}
  C_6^{\text{ind}} &= \frac{1}{2}q^2\alpha_{\text{q}}\,,
\end{align}
with the static electric quadrupole polarizability $\alpha_{\text{q}}$ of the
atom. The van der Waals type dispersion interactions are usually weaker, with
the leading-order term of Eq.~\eqref{Vdisp} accounting for dynamic
interactions due to instantaneous dipole-induced dipole moments of the ion and
the atom arising due to quantum fluctuations. Higher-order terms could be
added to Eq.~\eqref{Vdisp}, but since reaching the $s$-wave scattering regime
for \ce{Rb2+} is barely possible at all, truncating after the $R^{-6}$ is
usually sufficient.

The second term of Eq.~\eqref{VLR} is due to the indistinguishability of the
two limiting cases \ce{Rb}$+$\ce{Rb+} and vice versa. It is thus defined by
the energy splitting between the asymptotically degenerate gerade and ungerade
states and determines the cross section for resonant charge
transfer~\cite{Bardsley1975}. For alkali-metal \ce{X2+} systems, it generally
involves one active electron and if both ion and atom are in an $S$-state, it
takes the form, in \emph{atomic units}~\cite{Magnier1999,Cote2016}
\begin{subequations}
  \begin{eqnarray}
    V_{\text{exch}}(R) &=&
    \frac{V_{X{\,}^2\Sigma_g^+}(R)-V_{(1){\,}^2\Sigma_u^+}(R)}{2}\,,\label{VexchGeneral}\\
    &=& \frac{1}{2}AR^\alpha\mathrm{e}^{-\beta R}\left[1+\frac{B}{R}+\frac{C}{R^2}+\cdots\right]\label{Vexch}\,.
  \end{eqnarray}
\end{subequations}
In Ref.~\cite{Cote2016}, the parameters $\alpha$, $\beta$ and $B$ are related
by simple expressions to the ionization potential $I_{\ce{Rb}}$ of the \ce{Rb}
atom
\begin{align}
  \label{ParamsVexch}
  \beta &= \sqrt{2I_{\ce{Rb}}}\,,\quad \nu = \frac{1}{\beta}\,,\notag \\
  \alpha&=(2\nu -1)\,,\quad B =\nu^2\left(1-\frac{1}{2}\nu\right)\,.
\end{align}
The corresponding value for the ionization potential is taken from
Ref.~\cite{NISTRbAtomic,Lorenzen1983}, yielding
$I_{\ce{Rb}}^{\text{exp}}=0.15350655\,E_{\text{h}}$ (original measurement:
$I_{\ce{Rb}}^{\text{exp}}=33690.81\pm
0.01\,\mathrm{cm}^{-1}$). The parameter
$A$ is the normalization factor of the asymptotic wavefunction involved to
arrive at Eq.~\eqref{Vexch} and is given~\cite{Magnier1999,Smirnov2001} in
terms of the parameters of Eq.~\eqref{ParamsVexch}
\begin{align}
  \label{AmplitudeVexch}
  A &= -\frac{\beta^2(2\beta)^{2\nu}\mathrm{e}^{-\nu}}{\Gamma(\nu+1)\Gamma(\nu)}\,,
\end{align}
where $\Gamma(\cdot)$ denotes the Gamma function. The second-order expansion
coefficient $C$ of Eq.~\eqref{Vexch} may be extracted from fits to
\emph{ab-initio} results and is taken from Ref.~\cite{Cote2016} with
$C=-19.22$.

The characteristic ion-atom interaction length scale $R^*$ may be derived from
Eq.~\eqref{Vdisp}, yielding~\cite{RevModPhys.91.035001} $R^*=\sqrt{2\mu
  C_4\hbar^{-2}}$, which is in general at least one order of magnitude larger
than corresponding neutral atom-atom interactions. The corresponding
characteristic energy scale $E^*\propto (2\mu^2C_4)^{-1}$ is at least two
orders of magnitude smaller than the one for neutral atom-atom systems. Due to
the small reduced mass of \ce{Li2+}, this explains why it was possible to
reach the $s$-wave scattering regime in the experimental run described in
Ref.~\cite{Schmid2018}. For the \ce{Rb2+} system, the characteristic
interaction length scale is $R^*\approx 5000\,\mathrm{a}_0$, with the
respective $s$-wave scattering limit $E^*=k_{\text{B}}\times
79\,\mathrm{nK}$~\cite{Kleinbach2018}. This stringent temperature requirement
is one of the reasons why reaching the quantum collision regime for \ce{Rb2+}
is considerably more difficult as compared to neutral atom systems or to
\ce{Li2+}.

\subsection{\label{subsec:ConstrRb2p}Construction procedure}

The results from Sec.~\ref{subsec:Rb2p-basis-set-effects-HLC} indicate that overall the ROHF-CCSD(T) approach based on effective core potentials is already quite accurate. We will thus use it for the overall construction of the $X\,\tensor*[^2]{\Sigma}{_g^+}$ and $(1) \,\tensor*[^2]{\Sigma}{_u^+}$ potential energy curves and account for higher-level corrections by an appropriate rescaling.

\begin{figure}[tb]
  \centering
  \includegraphics[width=\columnwidth]{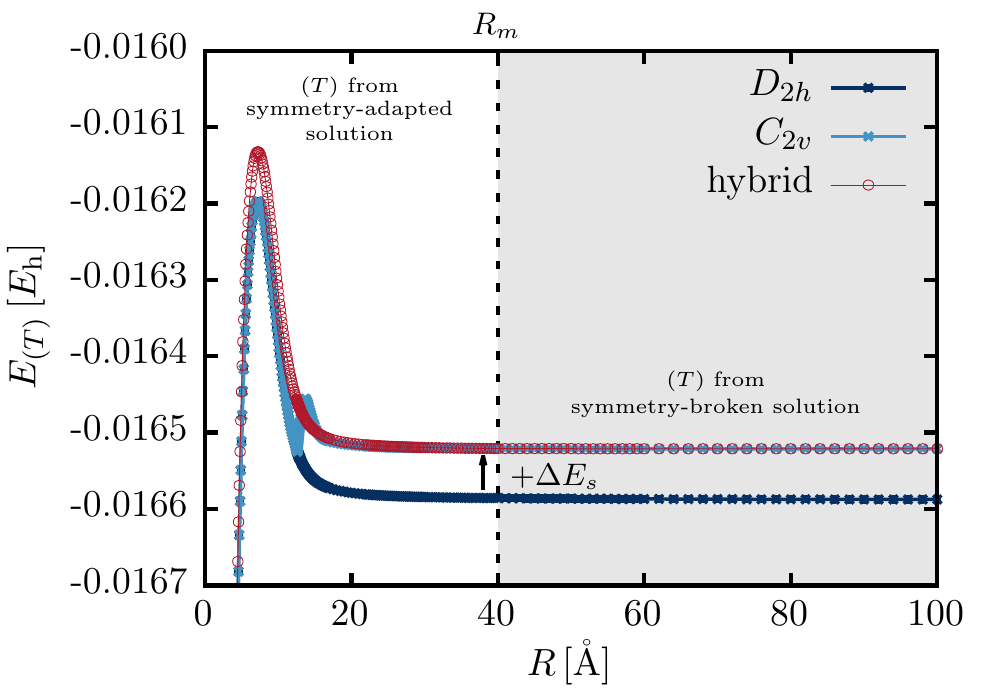}
  \caption{\label{fig:constr-proc-triples}Schematic visualization of the
    construction procedure illustrating how to merge symmetry-broken and
    symmetry-adapted (T) contributions to avoid the repulsive CCSD(T)
    long-range barrier and to consequently obtain a hybrid ROHF-CCSD(T) model.}
\end{figure}

Concerning the CCSD(T) approach for \ce{X2+} systems in general, our previous work~\cite{Schnabel2021} revealed some limitations, as already indicated at the start of this section. In Ref.~\cite{Schnabel2021}
we demonstrated that CCSD(T) leads to an unphysical
long-range barrier of the respective system, which is related to a
symmetry instability of the underlying Hartree-Fock mean-field
solution. However, our findings also suggested that using  (T)
corrections from symmetry-broken calculations for the long-range tail and properly merging these with symmetry-adapted
solutions for smaller internuclear distances may be a promising approach to
construct well-defined and physically meaningful global PECs for the
$X\,\tensor*[^2]{\Sigma}{_u^+}$ and $(1)\,\tensor*[^2]{\Sigma}{_u^+}$ states
of \ce{Rb2+}. The hybrid ROHF-CCSD(T) energies for both states are defined
as~\cite{Schnabel2021_PhD}
\begin{align}
  \label{hybr-CCSDpT}
  E_{\text{CCSD(T)}}^{\text{hybrid}}(R) &= E_{\text{ROHF}}^{D_{2h}}(R) +
  \Delta E_{\text{CCSD}}^{D_{2h}}(R) \notag \\ &+ \Delta E_{\text{(T)}}^{\text{hybrid}}(R)\,, 
\end{align}
with the hybrid (T) correction to model the long-range region given by
\begin{align}
  \label{hybr-T-contr}
   \Delta E_{\text{(T)}}^{\text{hybrid}}(R) &= \Delta
   E_{\text{(T)}}^{C_{2v}}(R)\Theta(R-R_m) \notag \\ &+\left[\Delta
     E_{\text{(T)}}^{D_{2h}}(R)+\left|\Delta E_s\right|\right]\Theta(R_m-R)\,,
\end{align}
where $\Theta(R)$ is the Heaviside function. The first two terms in
Eq.~\eqref{hybr-CCSDpT} denote the ROHF reference energy and the CCSD
correlation energy, respectively. The point group labels represent the
computational point groups and correspond to symmetry-adapted ($D_{2h}$) and
symmetry-broken ($C_{2v}$) solutions. Equation~\eqref{hybr-T-contr} formally
represents the use of (T) corrections from symmetry-broken calculations to
model the long-range tail and the proper merging to symmetry-adapted (T)
corrections at some merging point $R_m$ in the intermediate region to describe
the remaining part of the PEC. At $R_m$ the
symmetry-adapted values have to be shifted by the respective constant energy
difference $\left|\Delta E_s\right|$ to the symmetry-broken solution to obtain
continuous curves. This construction procedure is
schematically illustrated in Fig.~\ref{fig:constr-proc-triples}.
In total energies this shift amounts up to
$\approx\mathcal{O}(14\,\mathrm{cm}^{-1})$, as can be approximated by the
respective differences in Fig.~\ref{fig:constr-proc-triples}. For interaction
energies, in particular in the vicinity of the potentials equilibrium
$R=\SI{4.8}{\angstrom}$, the respective difference is in the order of
$\approx\mathcal{O}(0.5\,\mathrm{cm}^{-1})$. In general, the point $R_m$,
cf. Eq.~\eqref{hybr-T-contr}, enters as an additional free parameter that
might be adjusted so to optimally reproduce certain experimental findings
(e.g. spectroscopic constants, scattering length, vibrational levels, etc.).

We further note, that the small oscillation that occurs in
Fig.~\ref{fig:constr-proc-triples} for the symmetry-broken curve before it
collapses to the symmetry-adapted one is a consequence of the numerical
bistability of the symmetry-broken CCSD(T) solution in the region where it
cannot ``decide'' whether to collapse to the symmetry-adapted $\Sigma_g$ or
$\Sigma_u$ state; see Refs.~\cite{Schnabel2021,Schnabel2021_PhD} for details.

The $X\,\tensor*[^2]{\Sigma}{_g^+}$ and $(1)\,\tensor*[^2]{\Sigma}{_u^+}$
states of \ce{Rb2+} are asymptotically degenerate and thus the construction
procedure according to Eqs.~\eqref{hybr-CCSDpT} and~\eqref{hybr-T-contr} can
be applied to obtain physically correct PECs for both states. This implies
that in both cases $R_m$ has to be chosen equally, which is only justified if
$\Sigma_g$ and $\Sigma_u$ are reasonably well degenerate for a given merging
parameter $R_m$. Here, we assume that this is fulfilled if the difference
between the total symmetry-adapted ROHF-CCSD(T) energies is $\leq
10^{-8}\,\mathrm{E_h}$. Moreover, the choice of $R_m$ should be sufficiently
far off the repulsive long-range barrier occurring at
$R\approx\SI{100}{\angstrom}$~\cite{Schnabel2021}. These two conditions define
the restriction
\begin{align}
  \label{restr-Rm}
  27.0\,\si{\angstrom}\leq R_m < 50.0\,\si{\angstrom}\,.
\end{align}
We set this value to $R_m=\SI{40.0}{\angstrom}$
for the following discussion.

As shown in the previous section it is important to extrapolate the
ROHF-CCSD(T) results to the respective basis set limits. Again, we found an
approach based on Eqs.~\eqref{CBSHF} and~\eqref{BSECCSDptCorrEnerg} as the
best compromise to obtain the CBS values using the
UET17 basis sets. Effects beyond the CCSD(T)/ECP approach, as discussed in Sec.\ \ref{subsec:Rb2p-basis-set-effects-HLC},
are now simply included by rescaling the potential energy curves to match $D_e$ and $R_e$ of the theoretical best
estimates.

To generate PECs that can be used to study \ce{Rb+}$+$\ce{Rb} scattering
events it is inevitable to recover the correct long-range behavior according
to Eq.~\eqref{VLR}. This also involves ensuring that the rescaled \emph{ab-initio} data
reproduce the theoretically suggested exchange splitting between the gerade
and ungerade states as given by Eq.~\eqref{Vexch}. The following protocol is
used to incorporate the exchange interaction into the PEC construction
procedure~\cite{Schnabel2021_PhD}:
\begin{enumerate}
\item Compute the difference $\Delta(R)$ between the theoretically suggested
  exchange splitting $V_{\text{exch}}(R)$ of Eq.~\eqref{Vexch} [with
    corresponding parameters from Eqs.~\eqref{ParamsVexch}
    and~\eqref{AmplitudeVexch}] and the one
  resulting from the rescaled \emph{ab-initio} data $\widetilde{V}_{\text{exch}}$
  according to Eq.~\eqref{VexchGeneral}
  \begin{align}
    \Delta(R) &= V_{\text{exch}}(R) - \widetilde{V}_{\text{exch}}(R)
  \end{align}
\item Equation~\eqref{Vexch} is only valid for~\cite{Smirnov2001}
  \begin{align}
    R\beta\gg 1\qquad\text{and}\qquad R\beta^2\gg 1\,.
  \end{align}
  In the following we assume that this means that both products have to be at
  least one order of magnitude larger than one. This yields a validity interval
  $\mathbb{V}$ of internuclear distances $R$
  \begin{align}
    \label{VexchValidity}
    R\in\mathbb{V}=[40.0,\infty)\,\mathrm{a}_0\approx [21.0,\infty)\,\si{\angstrom}
  \end{align}
  within which the rescaled PECs have to be modified. Note that this
  is independent from the merging parameter $R_m$ as introduced above to join
  symmetry-broken and symmetry-adapted (T) solutions in the long-range
  tail.
\item The long-range parts $\widetilde{V}_{\text{LR}}$ of
  the $X\,\tensor*[^2]{\Sigma}{_g^+}$ and $(1)\,\tensor*[^2]{\Sigma}{_u^+}$
  states will reproduce $V_{\text{exch}}$ of Eq.~\eqref{Vexch} if
  they are \emph{additionally} modified according to
    \begin{align}
    \label{VexchModSigma-g}
    \widetilde{V}_{\mathrm{LR}}^{\Sigma_g^+}(R) &=
    \begin{cases}
      \widetilde{V}_{\mathrm{LR}}^{\Sigma_g^+}(R) +
      \Delta(R_v=40.0\,\mathrm{a}_0)\,,\quad R <
      40.0\,\mathrm{a}_0\\
      \widetilde{V}_{\mathrm{LR}}^{\Sigma_g^+}(R) + \Delta(R)\,,\quad R\geq 40.0\,\mathrm{a}_0
    \end{cases}
    \,,
  \end{align}
  and
  \begin{align}
    \label{VexchModSigma-u}
    \widetilde{V}_{\mathrm{LR}}^{\Sigma_u^+}(R) &=
    \begin{cases}
      \widetilde{V}_{\mathrm{LR}}^{\Sigma_u^+}(R) -
      \Delta(R_v=40.0\,\mathrm{a}_0)\,,\quad R <
      40.0\,\mathrm{a}_0\\
      \widetilde{V}_{\mathrm{LR}}^{\Sigma_u^+}(R) - \Delta(R)\,,\quad R\geq 40.0\,\mathrm{a}_0
    \end{cases}
    \,,
  \end{align}
  where $R_v$ labels the lower bound of the validity
  interval~\eqref{VexchValidity}.
\end{enumerate}
An overview of exchange splittings $\widetilde{V}_{\text{exch}}$ that result
from \emph{ab-initio} data and a corresponding comparison with the respective
theoretical curve according to Eq.~\eqref{Vexch} is given in the supplementary
material~\cite{Supplementary}. Moreover, it is shown that the exchange interaction is very
sensitive to the respective basis set leading to an interchange of the
$\Sigma_g$ and $\Sigma_u$ states; a problem that already occurs at the SCF
level.

\begin{figure}[tb]
  \centering
  \includegraphics[width=\columnwidth]{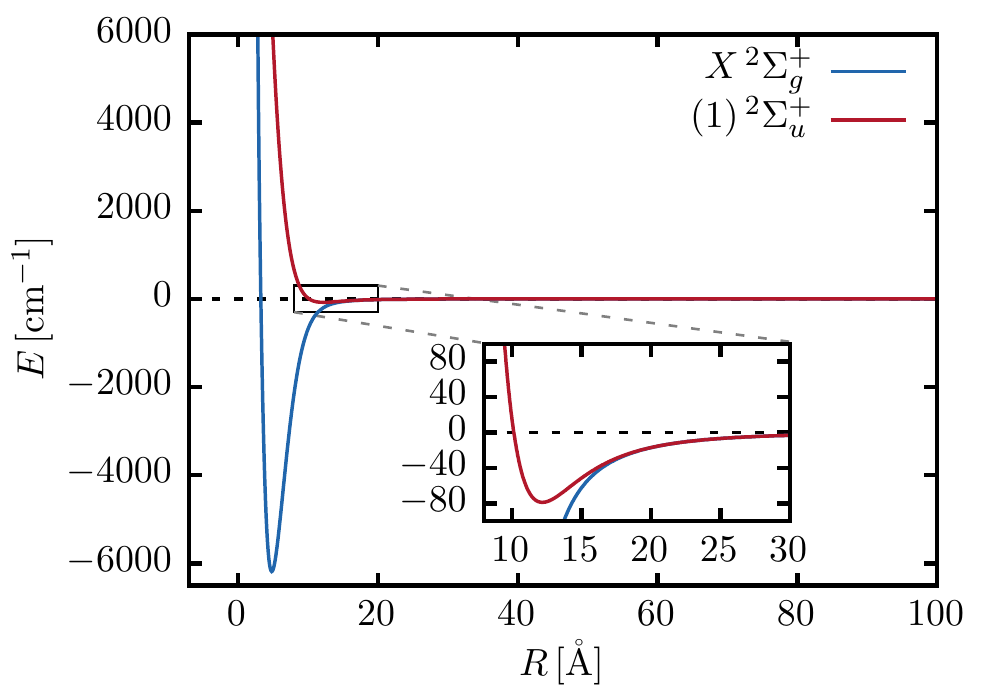}
  \caption{\label{fig:Rb2p-RPRKHS-PEC_Sg_Su}Potential energy curves of the
    $X\,\tensor*[^2]{\Sigma}{_g^+}$ and $(1)\,\tensor*[^2]{\Sigma}{_u^+}$
    states of \ce{Rb2+} resulting from
    the RP-RKHS interpolation method
    based on \emph{ab-initio} data obtained at hybrid ROHF-CCSD(T)/CBS/mod
    level of theory with the UET17 basis set. The inset
    shows the shallow potential well corresponding to the ungerade state.}
\end{figure}

Finally, the adapted \emph{ab-initio} data ($\equiv$ ``hybrid
ROHF-CCSD(T)/CBS/mod'' level of theory), i.e. those obtained by first
transforming according to the hybrid model of Eqs.~\eqref{hybr-CCSDpT}
and~\eqref{hybr-T-contr}, followed by a rescaling to the theoretical best estimates for 
$D_e$ and $R_e$ and then by subsequent modification to obtain the
correct exchange splitting using Eqs.~\eqref{VexchModSigma-g}
and~\eqref{VexchModSigma-u}, are passed to the one-dimensional RP-RKHS
interpolation
method~\cite{Ho1996,Hollebeek1999,Ho2000a,Ho2000b,Higgins2000,Unke2017,SoldanPESRb3}. This
gives analytical representations for the $X\,\tensor*[^2]{\Sigma}{_g^+}$ and
$(1)\,\tensor*[^2]{\Sigma}{_u^+}$ PECs and yields by construction the correct
leading-order multipole terms according to Eq.~\eqref{Vdisp}. It is important
to include a sufficient amount of training data in the region $R\in
[20.0,60.0]\,\si{\angstrom}$ to ensure precise interpolation of the exchange
interaction~\cite{Schnabel2021_PhD}. Beyond that region, the training data
should be chosen sparsely to ensure extrapolation after Eq.~\eqref{Vdisp}. The
corresponding induction coefficients follow from the experimentally measured
static electric dipole polarizability
$\alpha_{\text{d}}$~\cite{Schwerdtfeger2019,Maxwell2015} yielding [see
  Eq.~\eqref{C4ind}] $C_4^{\text{ind}}=2.751960345\cdot
10^{6}\,\mathrm{cm}^{-1}\si{\angstrom}^4$ and from taking the most recent
value, based on relativistic coupled-cluster calculations from
Ref.~\cite{Kaur2015}, for the static electric quadrupole polarizability
$\alpha_{\text{q}}$ yielding [see Eq.~\eqref{C6ind}]
$C_6^{\text{ind}}=0.156412961\cdot
10^8\,\mathrm{cm}^{-1}\si{\angstrom}^6$. The RP-RKHS parameters are chosen as
$n_{\ce{Rb2+}}=2$, $m_{\ce{Rb2+}}=1$, $s_{\ce{Rb2+}}=2$ and
$R_a=\SI{20.0}{\angstrom}$ for the $X\,\tensor*[^2]{\Sigma}{_g^+}$ state and
$R_a=\SI{85.0}{\angstrom}$ for the $(1)\,\tensor*[^2]{\Sigma}{_u^+}$
state. Furthermore, Eq.~\eqref{VSR} was imposed to provide a physically
meaningful short-range description. Again, the RP-RKHS parameters and the
short-range correction do not affect $D_e$ and $R_e$.
The resulting PECs can be generated using the \textsc{python} script and data
available from Ref.~\cite{RPRKHS_Rb2plus_Schnabel2021}, which are either based
on the aug-cc-pCV$n$Z-PP or the UET17 basis set families both
extrapolated to their respective CBS limit. Corresponding data were generated
with the hybrid ROHF-CCSD(T) approach according to Eqs.~\eqref{hybr-CCSDpT}
and~\eqref{hybr-T-contr} and modified to reproduce the correct exchange
interaction using Eqs.~\eqref{VexchModSigma-g}
and~\eqref{VexchModSigma-u}. This procedure yields the PECs depicted in
Fig.~\ref{fig:Rb2p-RPRKHS-PEC_Sg_Su}.

\subsection{\label{subsec:Results-Rb2p}Resulting potential energy curves}

As displayed in Fig.~\ref{fig:Rb2p-RPRKHS-PEC_Sg_Su}, the ground state of
gerade symmetry is much deeper compared to the asymptotically degenerate state
of ungerade symmetry, which only shows a rather shallow well at about
\SI{12}{\angstrom}. We note that the choice of the merging parameter $R_m$,
within the constraints defined by Eq.~\eqref{restr-Rm}, mainly affects the
long-range behavior of the respective PECs. It can thus be viewed as defining
a lower bound of the fitting range if the corresponding \emph{ab-initio} data
were used to extract higher-order induction and dispersion coefficients
(i.e. $C_6^{\text{disp}}$, $C_8^{\text{ind}}$, etc.). These higher-order
coefficients might improve the quality of the RP-RKHS interpolated PECs by
including them into the inherent extrapolation according to
Eq.~\eqref{Vdisp}. This in turn may be used to screen the sensitivity of
subsequent scattering calculations based on these RP-RKHS PECs that account
for such higher-order effects. A more detailed discussion on the extraction of
induction and dispersion coefficients from hybrid ROHF-CCSD(T)
\emph{ab-initio} data may be found in Ref.~\cite{Schnabel2021_PhD}.

As already discussed in Sec.~\ref{subsec:ConstrRb2p}, the choice of the
merging point $R_m$ and thus of $|\Delta E_s|$, cf. Eq.~\eqref{hybr-T-contr},
leaves $R_e$ effectively unchanged and alters $D_e$ by about
$0.5\,\mathrm{cm}^{-1}$.~\cite{Schnabel2021_PhD}. The corresponding effects that result from 
accounting for the theoretically suggested exchange interaction are also very small
and give rise to the tiny differences between the theoretical best estimate and the model potential for 
the binding energies $D_e$ as documented in Tab.~\ref{tab:SpectroscConstants_Sg-Su}.
The rovibrational ground state
energies, i.e. $(v,J)=(0,0)$ where $v$ and $J$
denote the vibrational and rotational quantum numbers, were extracted using
the \textsc{Level16} code~\cite{Level16} and the final RP-RKHS
potential energy curve. The results are summarized in Tab.~\ref{tab:SpectroscConstants_Sg-Su}. 

\begin{table*}[tb]
  \centering
  \caption{\label{tab:SpectroscConstants_Sg-Su}Overview of the theoretical best estimates and the
    binding energies and lowest rovibrational states realized in the fitted model potentials for the 
    $X\,\tensor*[^2]{\Sigma}{_g^+}$ and
    $(1)\,\tensor*[^2]{\Sigma}{_u^+}$ states of \ce{Rb2+} in comparison to available experiments and other theoretical work.
    }
  \begin{adjustbox}{max width=\textwidth}
    \begin{tabular}{l@{\hskip 20pt}c@{\hskip 35pt}c@{\hskip 35pt}c@{\hskip 35pt}c@{\hskip 35pt}c@{\hskip 35pt}c}
      \toprule\toprule
      \multirow{2}{*}{basis set} &
      \multicolumn{3}{c}{$X\,\tensor*[^2]{\Sigma}{_g^+}$} &
      \multicolumn{3}{c}{$(1)\,\tensor*[^2]{\Sigma}{_u^+}$} \\
      \cmidrule(r){2-4} \cmidrule(r){5-7} & 
      \multicolumn{1}{c}{$D_e\,[\mathrm{cm}^{-1}]$} & \multicolumn{1}{c}{$D_0\,[\mathrm{cm}^{-1}]$} &
      \multicolumn{1}{c}{$R_e\,[\si{\angstrom}]$} & \multicolumn{1}{c}{$D_e\,[\mathrm{cm}^{-1}]$} &
      \multicolumn{1}{c}{$D_0\,[\mathrm{cm}^{-1}]$} & \multicolumn{1}{c}{$R_e\,[\si{\angstrom}]$} \\
      \midrule
               TBE     &  6202 $\pm$ 30 &  & 4.796 $\pm$ 0.010 & 80.4 $\pm$ 9.0 & & 12.163 $\pm$ 0.020 \\
              {Model pot. (${}^{87}\ce{Rb2+}$)} & 6202.02 & 6179.07 & 4.796 &
              80.31 & 78.46 & 12.163 \\
              {Model pot. (${}^{85}\ce{Rb2+}$)} & 6202.02 & 6178.80 & 4.796 &
              80.31 & 78.44 & 12.163 \\
          \midrule
          {Experiments} & ~ & ~ & ~ & ~ & ~ & ~ \\
          \cite{Lee1965,Shafi1972} & 5888$\pm$484 & {--} & {--} & {--} & {--} & {--} \\
          \cite{Olsen1969} & {--} & {--} & 3.94 & {--} & {--} & {--}
          \\
          \cite{Wagner1985} & 6049$\pm$807 & {--} & {--} & {--} &
               {--} & {--} \\
          {\cite{Bellos2013} (${}^{85}\ce{Rb2+}$)} & {--} & $\geq 6307.5(6)$ & {--} & {--} & {--} &
               {--} \\
          \midrule
          {Theory (model potentials)} & ~ & ~ & ~ & ~ & ~ & ~\\
          \cite{Bellomonte1974} & 6936 & {--} & 4.445 & ~ & ~ & ~ \\
          \cite{Patil2000} & 6977 & {--} & 4.604 & ~ & ~ & ~ \\
          \cite{Aymar2003} & 5816 & {--} & 4.868 & {--} & {--} & {--} \\
          {Theory (lcECP$+$CPP)} & ~ & ~ & ~ & ~ & ~ & ~ \\
          \cite{Laszlo1982} & 6130 & {--} & 4.780 & ~ & ~ & ~ \\
          \cite{LargeCoreECP} & 6365 & {--} & 4.820 & ~ & ~ & ~ \\
          {\cite{Krauss1990} (compact effective potential)} & 6323 & {--} & 4.731 & ~ & ~ & ~ \\
          \cite{Jraij2003} & 6167 & {--} & 4.794 & 82 & {--} &
          12.1340 \\
          {\cite{Bellos2013} (${}^{85}\ce{Rb2+}$)} & {--} & 6200$\pm$120 &
          {--} & {--} & {--} & {--} \\
          \bottomrule\bottomrule
    \end{tabular}
  \end{adjustbox}
\end{table*}

Experimental data given in this table  refer to
the works already mentioned in the introduction.
The most recent experiment~\cite{Bellos2013} aimed at measuring the ionization
potential of ${}^{85}\mathrm{Rb}_2$ formed via photoassociation of ultracold
${}^{85}\mathrm{Rb}$ atoms, cf. Sec.~\ref{sec:Intro}. The molecules were subsequently excited by
single-photon UV transitions to states above the ionization threshold. This
approach yielded an upper limit for the ionization energy of
${}^{85}\mathrm{Rb}_2$ and simultaneously provided a lower bound for $D_0$ of
${}^{85}\mathrm{Rb}_2^+$; as reported in
Tab.~\ref{tab:SpectroscConstants_Sg-Su} with $D_0^{\text{exp}}\geq
6307.5\,\mathrm{cm}^{-1}$.

As indicated in Tab.~\ref{tab:SpectroscConstants_Sg-Su}, theoretical works on
\ce{Rb2+} found in the literature can be divided into two classes:
calculations that are based on model
potentials~\cite{Bellomonte1974,Patil2000,Aymar2003} or approaches using
large-core effective core potentials (lcECPs) with a core polarization
potential (CPP)~\cite{Laszlo1982,LargeCoreECP,Krauss1990,Jraij2003,Bellos2013}
to account for polarization effects of the core electrons. In the latter
approach spin-orbit effects can be incorporated into the pseudopotential, cf.,
e.g., Refs.~\cite{LargeCoreECP,Jraij2003}. The full valence configuration
interaction \emph{ab-initio} calculations that accompany
the experimental work in Ref.~\cite{Bellos2013} were based on a relativistic
non-empirical large-core pseudopotential with CPP. Spin-orbit effects were
modeled through a semi-empirical spin-orbit pseudopotential and taken into
account in the valence CI calculations via the CIPSO procedure
(configuration interaction with perturbation including spin-orbit
coupling)~\cite{Allouche2012}. As can be inferred from the numbers reported in
Ref.~\cite{Jraij2003}, the SO effects on the spectroscopic constants of the two lowest
states of \ce{Rb2+} are negligible.

In comparison to the most recent experimental reference value $D_0^{\text{exp}}\geq
6307.5\,\mathrm{cm}^{-1}$ (Ref.~\cite{Bellos2013}) the binding energy computed in this
work comes out too low by approximately $120\,\mathrm{cm}^{-1}$. While this is in general a satisfactory
agreement, the discrepancy is somewhat in conflict with the accuracy estimates given by
us previously, which amount to $\mathcal{O}(30\,\mathrm{cm}^{-1})$.

\begin{figure*}[tb]
  \centering
  \begin{minipage}[t]{.49\linewidth}
    \includegraphics[width=\columnwidth]{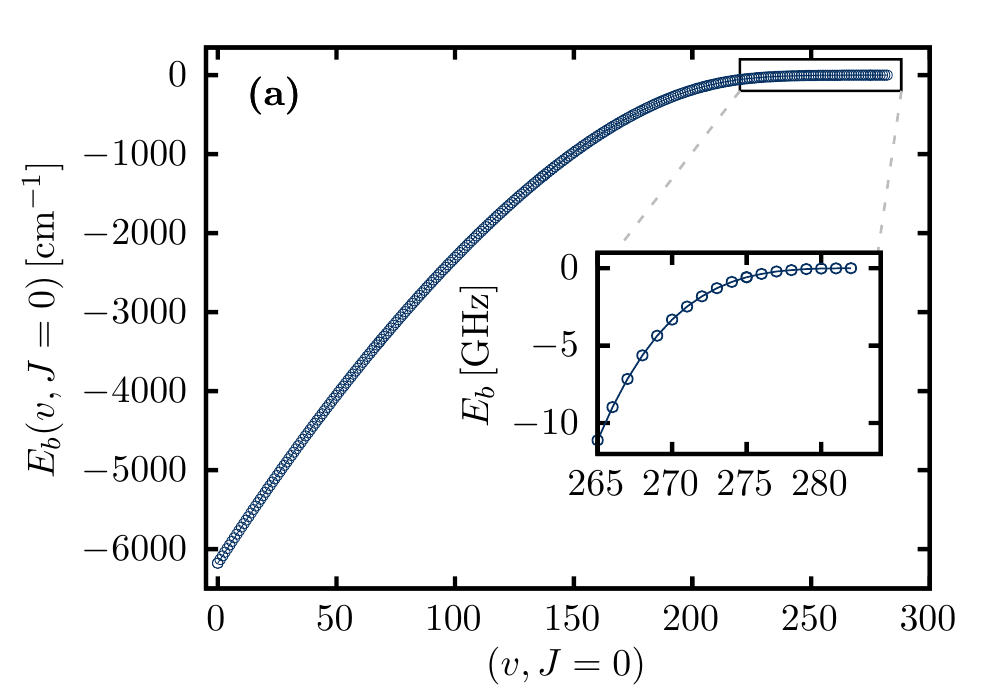}
  \end{minipage}
  \begin{minipage}[t]{.49\linewidth}
    \includegraphics[width=\columnwidth]{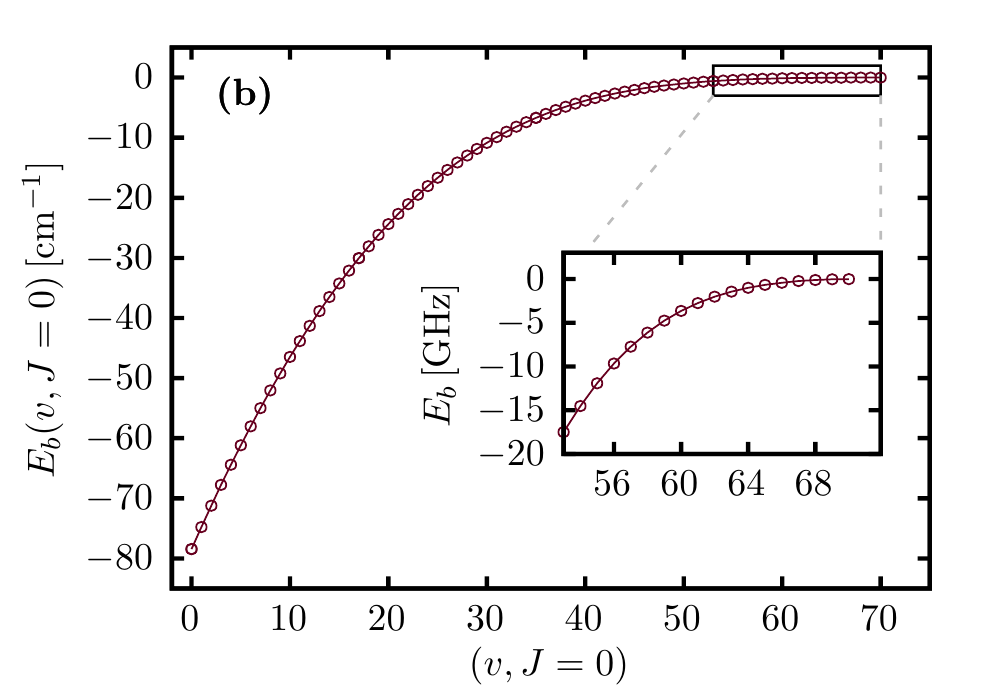}
  \end{minipage}
  \caption{\label{fig:rb2p_rovib}Computed vibrational levels ($J=0$) of (a) the
    $X\,\tensor*[^2]{\Sigma}{_g^+}$ state  and (b) the
    $(1)\,\tensor*[^2]{\Sigma}{_u^+}$ state.  The
    results correspond to the \ce{Rb2+} system which only contains the
    ${}^{85}\ce{Rb}$ isotope.}
\end{figure*}

\subsection{\label{subsec:Rb2pRovibStruct}Rovibrational structure}

The analysis of the rovibrational term values supported by the
$X\,\tensor*[^2]{\Sigma}{_g^+}$ and $(1)\,\tensor*[^2]{\Sigma}{_u^+}$ RP-RKHS
interpolated PECs based on hybrid ROHF-CCSD(T)/\-UET17(CBS)/\-mod
\emph{ab-initio} data implies the existence of approximately 280 and 70 vibrational levels ($J=0$),
respectively. These calculations were performed using the \textsc{Level16}
program~\cite{Level16} assuming that \ce{Rb2+} exclusively contains the
${}^{85}\mathrm{Rb}$ isotope. From a fit to the lower part of the spectrum (up to $v=50$ and $v=20$ for the two potentials) to
a Dunham expansion, the spectroscopic constants listed in Tab.~\ref{tab:vib-rot-consts} can be extracted. 
The corresponding values for $^{87}$\ce{Rb2+} are listed in the supplementary material~\cite{Supplementary}.
\begin{table}[tb]
  \centering
  \caption{\label{tab:vib-rot-consts}Spectroscopic constants of the $^{85}$\ce{Rb2+} states obtained by a fit to the lowest rovibrational states ($v\in\{0, 50\}$, $J\in\{0, 20\}$ for the $X\,\tensor*[^2]{\Sigma}{_g^+}$ state and $v\in\{0, 20\}$, $J\in\{0, 10\}$ for the $(1)\,\tensor*[^2]{\Sigma}{_u^+}$ state).}
  \begin{adjustbox}{max width=\textwidth}
    \begin{tabular}{lSS}
      \toprule\toprule
            {Parameter}   & \multicolumn{1}{c}{$X\,\tensor*[^2]{\Sigma}{_g^+}$} & \multicolumn{1}{c}{$(1)\,\tensor*[^2]{\Sigma}{_u^+}$} \\
          \midrule
            $\omega_e$ [cm$^{-1}$]      &  46.482   &  3.767     \\
            $\omega_ex_e$ [cm$^{-1}$]   &  8.05E-2  &  5.25E-02  \\
            $\omega_ey_e$ [cm$^{-1}$]	&  9.04E-05 &  2.46E-05  \\
            $\omega_ez_e$ [cm$^{-1}$]	& -6.19E-07 &  3.10E-06  \\
			$B_e$	[cm$^{-1}$]			&  1.72E-02 &  2.61E-03  \\
			$D_e$	[cm$^{-1}$]			& -8.94E-08 & -1.94E-06  \\
			$H_e$	[cm$^{-1}$]			& -2.90E-10 & -2.15E-08  \\
			$L_e$	[cm$^{-1}$]			& < 1E-12   &  8.80E-11  \\
			$\alpha_e$	[cm$^{-1}$]		&  3.96E-05 &  3.90E-05  \\
			$\beta_e$	[cm$^{-1}$]		&  4.35E-10 &  1.08E-08  \\
			$\gamma_e$	[cm$^{-1}$]		& -8.14E-09 & -1.45E-07  \\
              \bottomrule\bottomrule
    \end{tabular}
  \end{adjustbox}
\end{table}
The rovibrational structures of the two states for ($v,J=0$) are shown in
Figs.~\ref{fig:rb2p_rovib}~(a) and~(b). The level spacings of the deeply bound vibrational
states of the $X\,\tensor*[^2]{\Sigma}{_g^+}$ state amount to about
$46\,\mathrm{cm}^{-1}$, close to values reported from earlier computations \cite{Bellos2013}.

The $(1)\,\tensor*[^2]{\Sigma}{_u^+}$ potential is extremely shallow as reflected by $\omega_e\approx 3.8$ cm$^{-1}$.
Despite its well-depth of only $\approx 80\,\mathrm{cm}^{-1}$ it still can support more vibrational levels
than the $a\,\tensor*[^3]{\Sigma}{_u}$ state of \ce{Rb2} (cf. Tab.~S.V) with a
potential depth of $\approx 241\,\mathrm{cm}^{-1}$, for which only 41 bound states are found. This is a direct
consequence of the exceedingly large interaction length scale $R^*$ of
\ce{Rb2+}, which is at least one order of magnitude larger as compared to the
neutral species~\cite{RevModPhys.91.035001}.

All data generated by the \textsc{Level16} code can be found in the
supplementary material~\cite{Supplementary}, which also includes the corresponding input files for
technical details and to rerun our calculations. An analogous study based on
the aug-cc-pCV$n$Z-PP basis set series and referring to \ce{Rb2+} containing
merely the ${}^{87}\ce{Rb}$ isotope may be found in
Ref.~\cite{Schnabel2021_PhD}.

\section{\label{sec:Conclusion}Conclusions}
This work provides a protocol for computing highly accurate binding energies and accurate global PECs for the
ground state of \ce{Rb2+} within an additivity scheme based on
ROHF-CCSD(T) calculations. The approach circumvents our recently
revealed limitations of perturbative coupled-cluster approximations by using a
hybrid model with symmetry-broken (T) corrections describing the respective
long-range part of the PEC properly merged to symmetry-adapted solutions for
smaller internuclear distances. Furthermore, the construction procedure is
designed such that the PECs for the $X\,\tensor*[^2]{\Sigma}{_g^+}$ and
$(1)\,\tensor*[^2]{\Sigma}{_u^+}$ states of \ce{Rb2+} reproduce the physically
correct exchange splitting. This is particularly important when using the
corresponding potentials for highly accurate studies in the context of
ultracold chemistry, e.g. for scattering calculations. In this regard, we
moreover provide ready-to-use analytical PEC representations obtained within
the framework of RP-RKHS interpolation.

We benchmarked the accuracy of our computational method for ionization energies
of \ce{Rb} as well as for spectroscopic constants and vibrational levels of the
$a\,\tensor*[^3]{\Sigma}{_u^+}$ triplet ground state of \ce{Rb2}. In both cases
we obtained very good agreement with respective experimental findings, which confirm the
tight error bars estimated for the uncertainty of the computed values. While in
this respect including high-level correlation (HLC) contributions and
high-level relativistic corrections are essential for ionization energies, we
found that they are of much less importance for 
interaction energies. For both \ce{Rb2} and \ce{Rb2+}, HLC effects are in the order of
$\approx 2-3\,\mathrm{cm}^{-1}$ for the total binding energy.

Our computational approach gives a value of 6202$\pm$30~cm$^{-1}$ for the dissociation
energy of the $X\,\tensor*[^2]{\Sigma}{_g^+}$ state and 80.4$\pm$9.0~cm$^{-1}$ for the antibonding 
$(1)\,\tensor*[^2]{\Sigma}{_u^+}$ state. We also investigated the rovibrational structure of these
states based on the model potentials fitted to our computed values using the RP-RKHS approach.
In particular we get a lowest rovibrational state in the $X\,\tensor*[^2]{\Sigma}{_g^+}$ with a binding energy
of 6179~cm$^{-1}$. 
There is a residual deviation of approximately 130~cm$^{-1}$ from the most recent experimental estimate for the lower bound on the
binding energy of \ce{Rb2+}. This is a 
closer agreement as compared to other theoretical works on \ce{X2+} systems, with
$\ce{X}\in\lbrace\ce{Li},\ce{Na},\ce{K},\ce{Rb}\rbrace$. However, the discrepancy is significantly larger than the estimate for the residual uncertainty of our computations.
This will certainly require further theoretical and experimental investigations.

The data that support the findings of this study are available within the
article and its supplementary material~\cite{Supplementary}.

\begin{acknowledgments}
  J.S. and A.K.  acknowledge funding by
  IQ\textsuperscript{ST}. The research of IQ\textsuperscript{ST} is
  financially supported by the Ministry of Science, Research, and Arts
  Baden-W\"urttemberg. The authors furthermore acknowledge support by the state of Baden-W\"urttemberg through bwHPC
and the German Research Foundation (DFG) through grant no INST 40/575-1 FUGG (JUSTUS 2 cluster).
  The work at Johns Hopkins University was
    supported by the National Science Foundation under Grant No. PHY-2011794. 
\end{acknowledgments}

\bibliographystyle{apsrev4-1}
\bibliography{references_rb2plus}

\end{document}